# Agile meets Assessments: Case Study on how to do Agile Process Improvement in a Very Small Enterprise

Jakob Diebold[1], Philipp Diebold[2,3][0000-0002-3910-7898] and Arthur Vetter[4]

[1] store2be gmbh, Oranienstraße 185, Berlin, Germany
`jakob.diebold@store2be.com`
[2] Fraunhofer IESE, Fraunhofer-Platz 1, Kaiserslautern, Germany
`philipp.diebold@iese.fhg.de`
[2] Bagilstein GmbH, Elbestraße 40, Mainz, Germany
`philipp.diebold@bagilstein.de`
[4] Karlsruhe Institute of Technology, AIFB, Kaiserstrasse 89, Karlsruhe, Germany
`arthur.vetter@partner.kit.edu`

**Abstract.** Smaller software companies, such as start-ups do not often follow an explicit process, but rather develop in a more or less unstructured way. Especially when they grow or customer involvement increases. This development without any structured process results in problems. Thus, our objective was the improvement of the current development process of one software start-up by introducing appropriate agile practices and eliciting their effects. For this reason, we performed a pre and post process assessment using interviews. Based on the initial assessment, agile practices were selected and implemented. Finally, the post assessment and additional code metrics served as controlling mechanism to check whether weak points are addressed. The comparison of the two assessments showed that 13 ISO29110 base practices have been improved by the introduced eight agile practices. Thus, even more aspects have casually been improved than initially planned. Finally, the additional retrospective with company employees showed how the introduced agile practices positively influenced their work.

## 1 Introduction

Development processes have shown to be important in the area of software development, even when it appeared to be an engineering discipline [1]. In the area of development processes, the way of development tremendously changed with the publication of the agile manifesto [2] from a traditional plan-based way to a more agile way.

Regularly, Software Process Improvement (SPI) initiatives are often considered for the sake of addressing existing problems of improvement goals, such as increasing quality or shorten time-to-market [3]. Especially the two mentioned issues are crucial for Very Small Enterprises (VSE) or Small Medium-sized Enterprises (SME) due to the fact that they often focus on one single product that needs to have a high quality and always new features within updates.

Compared to larger companies, these smaller ones normally provide a higher flexibility on different levels, e.g. dealing with customer changes. However, they also often work in an informal way without a structured process. To provide these companies with structured processes, common software process models like SPICE (ISO 15504) or CMMI were tailored to their needs, resulting in the ISO29110 [5], a process assessment and improvement standard for VSEs. Similar to the mentioned standards [4], the ISO 29110 is presenting the "What" to do. Thus, it does not contradict with agile processes, covering the "How".

For this reason, our work combines both aspects, assessments with ISO 29110 as well as process improvement using agile practices. Our overall research goal is to identify which agile practices to select for improving the process and how do these changes influence process covered by assessments. Within a case study of a VSE, improvement suggestions, namely agile practices, are derived from the pain areas identified during the initial assessment. Based on a second assessment we were able to compare both and get a good impression of what the different agile practices improved or affected.

The remainder of this paper is structured as follows: Section 2 provides some background and related work. This is followed by the design of the case study (Section 3). The main section contains the results of both assessments, the implementation of the agile practices as well as the retrospectives (Section 4). Finally, we discuss the threats of this study (Section 5) and conclude the paper (Section 6).

## 2 Background & Related Work

**Agile.** Different studies [6][7] showed that agile development is the leading style of development approaches, especially in information systems. Even if Scrum is the dominating method within this development area, Diebold et al. [8] showed that it is common to not use any of the methods as given by books but adapting it to the specific (companies' or projects') context. For this reason, there is a little shift from the introduction of complete agile methods to smaller elements of the methods, the agile practices. Besides the usage of Scrum as common method, VersionOne [6] also provides a list of commonly top-used agile practices (they call them techniques) mentioned by their participants: Daily standup, Prioritized backlogs, Short iterations, Retrospectives, Iteration and Release planning, Unit testing, Team-based estimation, and Taskboard.

**ISO 29110.** The ISO 29110 was developed based on the ISO 12207 as well as CMMI and adapted to the needs of VSEs. This assessment standard is broken down into the common five parts: overview (part 1), framework and taxonomy (part 2), assessment guide (part 3), different profiles (part 4), and management guide for these profiles (part 5). Parts 3 and 5, which are also most important for our work, address the improvement of the company itself. Due to the context of VSEs, the standard only considers two processes in the process assessment model (PAM), "Project Management" (PM) and "Software Implementation" (SI). These processes are refined into base practices (BPs), similar to common assessment models, to evaluate the fulfillment.

Laporte et al. [9][10] performed several case studies in different VSEs with this PAM. Within the first study [9] all ISO 29110 work products and BPs were introduced, focusing on documentation. The second study [10] with two cases focused on documentation of the architecture and tests. Besides the mentioned case studies that deal specifically with ISO 29110, many case studies were performed by adapting the common standards for assessments of smaller companies [11][12].

**Agile in Start-Ups and VSE.** After the ISO29110 related work, we now describe studies dealing with (very) small enterprises and start-ups as a subset of them, similar to our case study. In a grounded theory study [13], Coleman and O'Connor explored how software development processes of start-ups look like. They identified that start-ups as well as agile methods focus on products and are considered for smaller teams so that they fit together. However, he identified that many agile methods could not be applied due to the fact of a very small number of developers (even lower than suggested by Scrum or so) [14]. Even if they are applied, it is impossible to use them as dogmatic as given by the book. Instead they have to be adapted to the company's context [8].

O'Donnell and Richardson [15] introduced practices from Extreme Programming (XP) in an Irish SME without adapting them to their specific context. They identified faster improvements and higher adaptability. Nonetheless, they also recognized that the small team-size, the customer being the project manager, and less documentation, results in problems, e.g. making and understanding decisions without the necessary documentation in their case. Finally, they conclude that it is necessary to analyze which elements are appropriate for the specific context before introduction.

## 3 Study Design

In general, our study is a case study according to [16][17] with interviews in an assessment-mode as data collection method. We first present the objective of this study with its research questions, followed by the study procedure. Finally, the company's context is roughly presented.

### 3.1 Research Questions

The overall objective of this study was the elicitation of the current state and its improvement. This is typical for case study and action research, inducing improvements.

This study goal was refined into the following three research questions (RQ):

**RQ1:** Which agile practices are appropriate for this specific VSE context?
**RQ2:** How does the usage of agile practices improve the development process according to common assessment standards, in this case ISO 29110?
**RQ3:** How do these improvements improve quality and time-to-market?

### 3.2 Study Procedure

The overall procedure of the case study included the following four steps: After the Pre-Assessment (beginning July 2016), the Agile Practices were identified and introduced (Mid of August 2016), and finally the Post-Assessment was conducted (End of October).

**Step 1 - Pre-Assessment:** Within this assessment, two interviews (each 90 minutes) and one self-assessment (by the first author, being one of the three company employees; also 90 minutes) were conducted according to the ISO 29110 PAM [5]. All authors iterative defined a semi-structured interview guideline following the PAM with its base practices of the two process areas. Before starting with the interview questions[1], the interview purpose was stated. For the analysis of the interviews, we used the commonly used **N**(one)**P**(artially)**L**(argly)**F**(ully)-scale (cf. Table 1), defined for different assessment approaches such as SCAMPI [18].

**Step 2 - Identification of appropriate agile practices:** First, we identified the improvement spots, based on the rating of the base practices in the pre-assessment, focusing on "not achieved" (N) and "partially achieved" (P) practices (see Section 4.1). This complies with the first step of the Agile Deployment Framework [19]. Based on Literature [6][7] and our experience, mainly coming from Fraunhofer IESE as applied research institute [8][20], the agile practices that might address the previous none or bad covered base practices, were selected (see Section 4.2 for more details). These ideas were discussed in the author-team before bringing them up as final suggestions to the development team of the company.

**Step 3 - Implementation of selected agile practices:** Based on [19], the next step was the preparation of the implementation of the agile practices. This included also necessary adaptions of these practices [8], which is important in our company's context due to the small size of the team (which is less than suggested by Scrum). Even if literature suggests to perform process improvements in a step-wise or iterative approach [20], in our case all agile practices were implemented together due to time limitations. Besides the agile practices and their adoptions, we also documented difficulties and problems that appeared during the introduction.

**Step 4 - Post-Assessment including retrospective:** At the end of the case study, we performed another assessment, similar to the initial one (two interviews and one self-assessment with the same participants). Thus, we were able to trace the change of BPs within the ISO 29110-PAM. Since we mapped the improvement actions, the introduction of the agile practice, to the BPs which we wanted to improve, we were able to check whether it worked or not. In addition to this assessment, a retrospective with all study participants was performed to identify further advantages and disadvantages of this process improvement. Finally, besides the subjective manner of the assessments, we were interested in checking the implemented improvements objective measures. This is especially interesting for RQ3 regarding the quality of the software.

### 3.3 Company context

The company studied within this case study is a software start-up located in Karlsruhe Germany. With being founded in 2015 by three students they are a very-small enterprise (VSE) according to the European commission. These three were the main employees (including the first authors of this paper) supported by up-to two additional working students. At the beginning, all employees were working as developers, with the time being, some focused more on sales/marketing, consulting, or customization of the product for specific customer needs. Regarding their process, they were aware of the fact of having almost none or no structured one, just coding some kind or refactoring due to architecture and design and final testing. Before starting with this improvement initiative, no dedicated Agile Practices were used in their company respectively the process.

---

[1] Interview questions can be found here: http://doi.org/10.13140/RG.2.2.17706.93120

Their product is a project monitoring software for very large companies, developing large and complex products. Thus, the software provides a hierarchical dashboard with a flexible configuration of the specific project and/or customer needs. As an Excel-Add In it is completely integrated in the Microsoft Office-environment, written in Visual-Basic.NET. The software code was managed by GitHub. As support ticketing system, a self-hosted instance of osTicket was used.

## 4     Study Results

Within this section, we present the results of the overall case study, structured according to the parts of the presented procedure.

### 4.1     Pre Assessment

The pre and post assessment results are structured according to the two ISO 29110 processes. We evaluated each BP with the NPLF-scale of the SCAMPI approach [18].

**Project Management.** This process of the ISO 29110 includes 27 BPs that need to be checked for creating the companies' profiles. The overall objective of this process is the systematic project management of all software implementation activities during the life-cycle. All the ratings of the single BPs are presented in Table 1 (row 2) with the NPLF-scale. In the following paragraphs, we are going to provide some example descriptions of BPs, their ratings as well as reasons for the ratings.

BP1 of Project Management deals with the "Review of the statement of work". In our case, the company has no statement of work. Instead, it only has the customer offer, including the customers' requirements. This offer is informally and none-structurally reviewed. Thus, our rating resulted in a "partially". The second one (BP2) "Define with the customer the delivery instructions of each one of the deliverables specified in the statement of work" is "largely achieved" because the deliverables are included in the initial offer but not in a separate document for traceability or something else. The "Identification of tasks […] to produce the deliverables specific in the statement of work" (BP3) is done using either issues in GIT or the tool Asana (for changes that are not directly software-related). Since not all issues are specified on the same level of detail, it is "largely achieved". Since at this point in time, no "estimation of the duration to perform each task" (BP4) is performed, this BPs is obviously "not achieved".

Other examples are BP12-14 which all deal with the project plan as work product, from describing, verifying, to review. Since the company does not have a formal project plan, we need to rate all these three as "not achieved". This is exactly the opposite compared to BP20-21, which are both dealing with backup aspects and are both rated as "fully achieved". The company has a code-repository on GIT as well as a network drive, which both include an automatic backup mechanism.

**Table 1.** Results of pre- (row 2 "PM / 1" and 4 "SI / 1") and post-assessments (row 3 "PM / 2" and 5 "SI / 2") (N = not achieved, P = partially achieved, L = largely achieved, F = fully achieved according to the SCAMPI approach [18])

| Process / Assessment | BP1 | BP2 | BP3 | BP4 | BP5 | BP6 | BP7 | BP8 | BP9 | BP10 | BP11 | BP12 | BP13 | BP14 | BP15 | BP16 | BP17 | BP18 | BP19 | BP20 | BP21 | BP22 | BP23 | BP24 | BP25 | BP26 | BP27 |
|---|---|---|---|---|---|---|---|---|---|---|---|---|---|---|---|---|---|---|---|---|---|---|---|---|---|---|---|
| PM / 1 | P | L | L | N | P | L | N | P | P | P | N | N | N | N | F | P | P | L | F | F | F | P | L | P | P | F | L |
| PM / 2 | P | L | L | F | P | L | P | F | P | P | P | L | N | N | F | L | L | L | F | F | F | P | L | P | P | F | - |
| SI / 1 | P | N | L | L | P | P | L | N | N | P | L | P | F | N | L | L | L | L | N | L | L | N | F | | | | |
| SI / 2 | F | F | L | F | P | L | L | P | N | P | L | L | F | N | L | L | L | L | N | L | L | L | F | | | | |

**Software Implementation.** Compared to Project Management, this process only implements 23 BPs for the company's profile. This process targets the systematic performance of analysis, construction, integration, and testing activities of a new or changed software product. All the ratings of the single BPs are presented in Table 1 (row 4) with the NPLF-scale. In the following paragraphs, we are going to provide some example descriptions of BPs, their ratings as well as reasons leading to the ratings.

The first three BPs within the Software Implementation deal with the requirements, from documenting (BP1), verifying (BP2), to validation (BP3). For the first one, it is necessary to use all possible information sources and cover the feasibility as well as scope. Since all these aspects should be covered in a document, which is only in a task in GIT in the company's case (also BP4), we rated BP1 with "partially achieved". Most often the degree of detail is varying and the details are only known to the owner of the requirement, which increases communication. Since no review of the requirements is conducted, BP2 is easily rated as "not achieved". The validation is covering the customers' expectations. Either most requirements come directly from the customer and are thus later validated or are discussed within the team of the company's founders, which also communicates with the customers. Thus, a "largely achieve" is feasible.

Within the area of testing, ISO 29110 requires "Design or update unit test cases and apply them to verify that the Software Components implements the detailed part of the Software Design" (BP12). In our case the company does not have a lot of module tests so only very few modules are completely tested (2.99% of the lines of code, cf. Table 4). For this reason, we need to rate it as low "partially achieved". Following the created tests, it is necessary to "correct the defects found until successful unit test" (BP13). In the cases where the tests are existing in our case, this is "fully achieved". We decided to rate it like that due to the fact that the missing tests were already rated down before.

**Overall.** When considering the overview of the first assessment, it shows some very good covered areas as well as some gaps. Out of the 50 BPs of the ISO 29110, 12 (24%) were rated as "not achieved", 15 (30%) as "partially achieved", 16 (32%) as "largely achieved", and 7 (14%) as "fully achieved".

### 4.2 SPI using Agile Practices (RQ1)

After the pre assessment, we could identify a set of ISO 29110 BPs which we wanted to improve with agile practices. For this, we first need to find the appropriate agile practices that would help in improving the identified issues. Thus, we used common literature about existing agile practices and went into some details for those that seem to be helpful. We categorized the BPs that might be addressed with one or a set of agile practices, verified this, and named these categories according to the appropriate ISO 12207 processes (cf. Table 2, column 1 and 2). In addition to the agile practices, we also analyzed who is involved in the processes that are impacted by these improvements as well as how the agile practices need to be adapted to the specific company's context. Most of the adaptions, that will be presented when going through the different improvement areas, appeared during the implementation or usage of them.

**Table 2.** Identified improvement BPs, mapping to ISO 12207 processes (as improvement areas) and solutions

| ISO 12207 processes | ISO 29110 BPs | Addressing Agile Practices |
|---|---|---|
| Project Planning | PM.BP 4, 7, 11-14, SI.BP1,2 | User Stories, Backlog, Planning Poker with Story Points |
| Project Assessment and Control | PM.BP 16, 22 | Burn-Down-Charts |
| SW Construction | SI.BP 11 | Coding Standards |
| SW Qualification Testing | SI.BP 5, 10, 12 | Test-driven Development |
| SW Implementation | SI.BP 12, 17 | Definition-of-Done |
| SW Architectural Design and SW Documentation Management | SI.BP 9, 14, 19 | User Stories and their connections |

**Project Planning.** The major findings within this area are the lack of a project plan as well as managing requirements. These issues directly affect several BPs within both processes. Since we focus more on the engineering part, we identified the practice of User Stories [21] for managing requirements. This is also an ideal element for starting some kind of planning, per iteration as well as long-term. Furthermore these stories were arranged in a product backlog in which they were prioritized together with the customer. For a better effort estimation, which is strongly connected with the planning, we introduced the concept of story points [22] together with the planning poker practice [23]. This seems to be appropriate due to the fact that you do not need to think in concrete timings, e.g. hours or days, but gives a rather abstract comparison-mechanism.

We defined the user story to (1) need to bring value, (2) formulated actively, (3) allocated to a user group (as specific as possible), and (4) self-contained. Since the case company did not make the benefits/value explicit, following template was used (similar to common user story templates): *As a <user type>, I want to <function>*. An example is: *As a meeting organizer, I am able to allocate a user group to the action management.*

The backlog was created based on the set of user stories, written on post-its and stuck on the office wall. The final backlog also included some elements from story mapping since categories and clusters were included in a second dimension (besides the priority). After defining the initial set of stories, the team started the estimation with story points. They considered half a working day as one story point and not the recommended one day [21], due to their part-time development. Planning Poker was performed as stated in common guidelines [23].

During the implementation and usage of these agile practices belonging to project planning, no adjustments were necessary, so that the team kept them from the beginning of the case study up to the end and further.

**Project Assessment and Control.** The identified main issue here was the missing transparency of the project status during the development. Therefore, we selected the burn-down-chart as the appropriate agile practice to improve the project monitoring. Together with the Definition-of-Done (which we describe later in detail) we wanted the chart to show the already "done" stories and story points as well as the still open ones.

During the usage of this concept, the team recognized a distortion of the performed work. This is due to the necessarily performed reviews and updated regression-test list. Since these two activities were often forgotten, it resulted in less "done" stories during the sprint and, instead, many at the end. Additionally, this also has the effect of conducting a lot of reviews in a short time before finishing the sprint, product, or release. This results in less benefits of the burn-chart, such that often the team members did not update it. To overcome this problem, the team decided to conduct reviews of the already completely implemented user stories regularly (once a week). For that reason a specific column "to review" was integrated in their already used Github-Kanban-Board. Finally, they also found a mechanism within their analog backlog to order the user stories.

**SW Construction.** Even if the assessment did not show a large amount of improvement potential, we decided to introduce coding standards for new or refactored code. They focus on guidelines how the code should be structured and look like. This practice should simplify the communication, due to a common understanding [24], and further influence the (collective) ownership of the code [25].

To not burden the developers, we tried to keep the coding guidelines as small and easy as possible. The concrete conventions, which were defined by the team: (1) Variables are written in CamelCase; (2) Variables are declared at the beginning of a method (if possible); (3) Methods are named on the schema: verbNoun(Adjective); (4) Each method is commented with an XML-comment.

During the definition and usage of the coding guidelines, the necessity of some aspects was questioned, e.g. the XML-comments that take a lot of effort. The major problem that appeared with the practices was falling into oblivion, because they needed to be removed from the office whiteboard. To overcome this, the team made the guidelines visible on a project-wiki page and additionally pinned them printed on an office wall.

**SW Qualification Testing.** The major issue regarding testing was the lack of existing (module or unit) tests. To not just start implementing more of these tests, but using an agile practice, we decided to implement test-driven development (TDD), another important practice from XP. We and the team were aware of the fact that this practice is one of the most effort-intensive practices, since it needs to be established and often results in an intervention of the current development procedure. For this reason, it was decided to use this new way of testing and development only for new features and not for changes, which are normally quite small in their case.

During performing TDD a very specific problem appeared with their software product: Since it is an Excel-Add In which interacts with a lot of third-party products (mainly from MS Office), the testability of some modules or classes is very hard. After this finding, the team tried to decouple the single internal functionality from the interfaces. So, that it is easy to implement test at least for these internal functionalities.

**SW Implementation.** Besides the previous mentioned quality assurance activities, like testing and reviews, other aspects need to be considered such that a specific requirement or user story is finished. To overcome this issue and establish the things that need to be done, the Definition-of-Done (DoD) was selected to be introduced as agile practice. Since we already mentioned the specification of requirements as user stories, the defined list of criteria is applicable to every story: (1) Code is implemented, working, commented, and refactored; (2) Code is complying with coding guidelines; (3) Code is reviewed by another developer (4-eye principle); (4) Module-Tests are implemented (if possible), (5) UI-test with a set of specified inputs; (6) (New) functionality is part of the test-checklist considered for regression-testing; (7) All used Excel-versions are tested

One issue that appeared during the usage of the DoD was already mentioned in the project monitoring: Stories without the code review are not "done". Furthermore, the maintenance of the test-checklist for the regression testing was not updated for every story. The team put that down to the fact that the DoD was not present enough to all the developers, similar to the Coding guidelines. This was resolved similar to the Coding guidelines with a wiki-page as well as a printout in the office.

**SW Architectural Design & SW Documentation Management.** The main aspect in this area that appeared in the assessment was the missing traceability record. To overcome this missing piece of documentation between the parts in the software (e.g. module or class) and the requirements, the idea was to establish a reference to the specific user story in the code.

During the implementation of this, the team recognized that this is not realizable due to the fact that modules could not be referenced one-by-one to a single user story. Further they assume that it would be to effort intensive to reference all connected stories.

### 4.3  Post Assessment

Now we are going to present the second assessment results:

**Project Management.** Similar to the first assessment, Table 1 (row 3) with the NPLF-scale presents the results for all BPs of this ISO 29110 process. We are not providing the reasons behind all ratings, but rather on some examples that could be later used for the comparison of the first and second assessment.

The "assignment of roles and responsibilities" (BP6) is done implicitly for all employees (between the co-founders) because they strictly clarified their responsibilities within the team. Since this makes project-specific roles obsolete, it was rated as "largely achieved". The following two BPs deal with some effort estimation aspects, first "assigning start and completion date […] in order to create the Schedule of the Project

Tasks" (BP7) and second "Calculate and document the project Estimated Effort" (BP8). With regard to the first one, the company did not consider the two dates and thus cannot create a schedule out of it. Nevertheless, the estimated duration of performing the tasks and the priorities give a rough indication. This duration also can be used for the other BP, due to the fact the overall project effort /duration can be calculated. Thus, we are ending in the first case with a "partial" and in the second with a "fully achievement". Similar to the first assessment, there are no changes with regard to the project plan such that all BPs connected with this (BP13, 14) need to be rated with "not achieved".

**Software Implementation.** Similar to the first assessment, Table 1 (row 5) with the NPLF-scale presents the results for the BPs. We are not providing the reasons behind all ratings, but rather on some examples later used for the comparison.

Through the usage of GIT as version and configuration system, "incorporate the requirements specification to the software configuration" (BP4) is "fully achieved". For the "Document or update the Software Design" (BP5) the developer of a story suggests (design) solutions, of which one is selected by the team. Since the decision is not (formally) documented, we could only rate it with "partially". Nonetheless, the "Verify and obtain approval of the Software Design" on technical level is performed during the implementation. Furthermore, the code review (included in the DoD) does also consider the design. Thus, this BP is rated as "largely achieved".

Since no traceability record is created (BP9), similar to the pre-assessment, it is rated as "not achieved". The functionality behind the traceability record is covered to some extent by the management of branches and pull-requests in the version control system, including the unit test in the repository. But since there is no document for the design BP10 is rated as "partially achieved".

**Overall.** When considering the overall results of this second assessment, it shows that the company is on a good way. Out of the 50 BPs, 5 were rated as "not achieved", 12 as "partially achieved", 20 as "largely achieved", and 12 as "fully achieved". With these new results, the company achieved quite good improvements (which will be stated and discussed in the next sub-section) and learned a lot.

### 4.4 Comparison of both Assessments (RQ2)

On the high-level view, the integration of the agile practices could show improvements: Out of the 50 of the ISO 29110, 26% were rated better in the second one (13 BPs, cf. Table 3): 7 increased by 1 level (14%), 4 increased by 2 levels (8%, PM.BP8 & 12, SI.BP1 & 22), 2 increased by 3 levels (4% PM.BP4, SI.BP2). None got worse (only one could not be assessed in the second case) that means ¾ stayed the same.

When considering our processes (areas) from ISO 12207 we could see most improvements within project planning with six BPs from PM and three from SI. This is followed by Software Implementation with two SI BPs as well as Monitoring and Testing each with one. Five out of these 13 improved BPs were not expected (cf. Table 3, column 3) because of positive side-effects of the agile practices, such as User stories that provide input, help and support to many BPs. But besides that, we also had ten BPs which should be improved through our agile practices, where we could not achieve any changes (cf. Table 3, column 4).

**Table 3.** Improved ISO 29110 BPs after the introduction of the agile practices

| ISO 12207 processes | Expected, improved | Unexpected, improved | Expected, unimproved |
|---|---|---|---|
| Project Planning | PM: 4, 7, 11, 12; SI: 1, 2 | PM: 8, 17; SI: 4 | PM: 13, 14 |
| Project Asses.& Control | PM: 16 | | PM: 22 |
| SW Construction | | | SI: 11 |
| SW Qual. Testing | SI: 12 | | SI: 5, 10 |
| SW Implementation | | SI: 6, 22 | SI:17 |
| SW Arch. Design and SW Docu. | | | SI: 9, 14, 19 |

The improvements on project planning result on the one hand from user stories with a story board and backlog and on the other hand from planning poker with story points for estimation (PM.BP4). PM.BP7 and 8 are based on the combination of both. Within PM.BP11 and 12 the board with the prioritized stories replaces the functions of a project plan. Even if it is no formal project plan, it enables easier changes in the project progression (PM.BP17). The usage of the stories template also standardizes the management of requirements for the implementation (SI.BP1, 2, 4). Within the monitoring area, the burn charts could improve PM.BP16 especially through the visualization. The testing especially improved through the higher importance of creating unit tests in the practice TDD (SI.BP12). Even if not initially expected, the DoD influenced two BPs (SI.BP6, 22) through the manifestation of some activities, e.g. code reviews.

As seen in Table 3, some BPs could not be improved even if planned. Even if the user stories improved the project planning, they could not replace a formal project plan, which is the reason for not improving all expected ones. For the monitoring, there is not that much improvement since the team did not consequently track their effort so that a comparison is impossible. For some of the practices the "formal" documentation is missing (SI.BP5) and this also results in none-improvements in others (SI.BP10). In the architecture and documentation area, as mentioned during the introduction of the agile practice, it was not considered as useful to create a specific document.

## 4.5   Retrospective (RQ3)

We additionally decided to perform a retrospective with all the founders of the company regarding our improvement initiative with the two assessment. Thereby, we could identify the following positive as well as negative aspects.

As main positive aspect, the participants mentioned planning, effort estimation, and identification of dependencies. In general, the people internalized the importance of (development) processes and mentioned that now they are "really" performing the activities correct, e.g. managing requirements. Highlighted was the usage of user stories due to a clear definition and scoping of requirements resulting a better overview. Even if we believe, that the customer value of the single functionalities could be stated within the user stories template, they mentioned that just through the usage of their user story template, the connection to the customers' value improved.

The major learning of the improvement initiative was how companies could benefit from structured processes, such as some internal standards, e.g. programming style-guides. Furthermore, they could recommend others, to use all the practices which they implemented during the case study.

Besides these benefits, also some negative issues were mentioned, which were not that generic, as the positive ones, but rather specific for some practices. For example, the burn-down chart did not bring the expected results, due to the realization and team size. Furthermore, the temporarily absence of the visual representation of the coding guideline or DoD (in the office) showed a disadvantage. This was recognized quite early and resulted directly in an improvement.

### 4.6 Measurement (RQ3)

We further compared some code metrics before and after the introduction of the agile practices. They were collected within their common IDE using Add-Ons. We used these metrics for a better objective view to evaluate whether the quality of the software product improved, compared to the previous subjective results. We selected seven common code metrics (cf. Table 4) due to the suggestion of IESE-colleagues working on quality modelling and measurement:

Table 4. Comparison of Pre- and Post-Assessment code metrics

| Metric | Pre- & | Post-Assessment |
|---|---|---|
| Test coverage[2] | 2.99 % | 5.21 % |
| Comments / code | 6.2 % | 5.3 % |
| Documentation[3] | 7.2 % | 7.8 % |
| Lines / file | 183.77 | 173.60 |
| Methods / class | 13.57 | 13.19 |
| Average complexity [26] | 2.53 | 2.47 |
| Max. complexity [26] | 42 | 39 |

With the test coverage, as first metric, that increased by more than 2% for all lines of code (cf. Table 4, row 2), the objective measure confirms the subjective feeling that test-driven development (in our case study only for new features) increases the coverage of the tested code. Since this should probably result in finding bugs and improving the product, the quality increases.

Compared to all other metrics, the comments are the only one that got worse (cf. Table 4, row 3). The only reason for explaining we could imagine are the introduced coding conventions/style guides as well as lower (average) size of methods such that things do not need to be commented any more due to meaningful names. Contrary to that, the documentation within the code increased (cf. Table 4, row 4). Which was part of the coding conventions which was checked within code reviews.

---

[2] Percentage of lines of code executed by all unit tests
[3] Percentage of specific documentation comment lines

Due to the introduction of the coding guidelines, that include some conventions of file, class, and methods lengths, the lines per file as well as the method per class ration could be decreased (cf. Table 4, row 5 & 6). Furthermore, several rations of the different metrics were also considered. Especially the lines per file could be significantly reduced by 10 lines on average.

Finally, the last group of metrics covers complexity. McConnell's complexity metrics [26] showed an improvement (cf. Table 4, row 7 & 8). The average complexity could be improved slightly, which is the reason due to a large number of untouched code between the two assessments. Nonetheless, the maximum complexity, measured as possible paths through one methods, could be decreased.

Even if the time interval between the two measurements was quite short, except one metric, all showed an improvement; some more some less. Nonetheless, it is hard to generalize an improvement of the quality from these measures in this short interval.

## 5      Threats to Validity

Within this section we are going to discuss the threats regarding the validity of our study results according to [28] and how we tried to deal with them.

First of all, we are aware of the fact that our case company is not representative for all VSEs and much less for SMEs, mainly due to the fact that they are 1.5 years old and working with a small team. This also influences their daily work in the way, that they could not spend the same amount of time in the (further) development of their product every day or even over weeks, because of other activities such as sales, marketing, etc.

Regarding the case study execution, the time interval of four months between the two assessments and measurements were fixed by the company because of their product or customer times. For implementing and adapting the agile practices as well as obtaining an effect, it is a quite short interval. Even if we would have liked to have a longer time frame in between, we believe that our results indicate at least a trend. Further, we hope to conduct another measurement (including assessment) at a later point in time. Additionally, and also caused by the short timing, we needed to introduce the agile practices for the improvement all in parallel as big-bang and could not perform a step-by-step introduction which could be more beneficial for the company [20] as well as for research for measuring the effects of the single agile practices. This makes it also harder to identify the impacts, benefits, and drawbacks of the singe agile practices.

Furthermore, with performing assessments with interviews, we are facing common threats to validity that are common to these studies. Conducting interviews, always includes some possible bias from the interviewer, especially in our case were the interviewer was one of the company's founders. We tried to minimize this threat by using the ISO 29110 BPs as a standardized (and externally given) interview guideline. But we only focused on the BPs of the ISO and thus only implicitly considered the mentioned work products. The interviews were conducted with all founders (being the em-ployees) of the company to overcome the threat of having subjective individual viewpoints. Finally, the notes which we extracted from the audio tapes, were provided to the interview participants for a review, which increases the objectivity of the results. Nevertheless, all interviews or assessments done with interviews are at least to some extend subjective because of the personal skills and interpretations of the assessor. Asking for evidences in an assessment in the only way of trying to get some more objective results or statements on that.

For the analysis of both assessment results, besides the main author another researchers with assessment experience reviewed the results, provided input, and within a final discussion we ended up with the above rating. Similar for the selection of the appropriate agile practices, which was discussed among all authors.

Ending to the threats to validity, some more information on the possibility of replicating this study in a different context. For sure this is only possible in SME or even VSE for which this specific ISO-standard is developed for. Using the created interview guidelines that are based on the PAM with its base practices (BR) the replication of the interview can be done quite easily. The hardest part for a replication is the identification of the agile practices as improvement suggestions. This is the case because it requires experiences in agile development with knowing the single practices (even if having and using an existing experience base) as well as understanding the given context of the company. With that knowledge / experience and the information given above on the procedure the study can be replicated.

# 6  Conclusions

Within this paper we describe our case study in a software start-up. It comprises two ISO 29110-assessments with the implementation of agile practices in between as process improvement. Based on the two assessments with the same interview participants in both assessments, a good pre- and post-comparison was possible. This was supplemented by a retrospective and comparing objective measures of the product quality.

Based on all the results, the introduction and implementation of the eight agile practices (RQ1) showed significant improvements in the daily development of the case company. From the 50 base practices of the selected assessment standard, 13 were improved through the introduction of the agile practices (RQ2), whereas eight of them were expected to be improved and five additional ones improved. Based on the assessments and the retrospectives, user stories seem to be the most powerful agile practice in this case study. Nonetheless, we also expected ten to be improved that did not change, which we trace back to several aspects, e.g. missing agile culture/mind set, missing documentation-awareness in agile. But overall, the results were very positive.

Besides the concrete results presented in the previous paragraphs, we learned a lot during this study and would like to share these lessons such that other companies could benefit from:

- A pre-post-comparison is very helpful for evaluating the effects of the implemented (process) improvements
- External interviewers are better to provide objective results

- Some agile practices are easier to be established than others, for example UserStories can be integrated independent of most context issues. Whereas other practices, especially meetings, have the problem that they involve several people
- Experience in agile is necessary to suggest (appropriate) agile practices especially to fill the identified gaps from the initial analysis
- Often the practices do have more effects than expected, especially when considering some of them in combination

Even if the study gives a good idea of how software process improvement works by integration in agile practices, we have some ideas of how to further proceed within this study and topic in general. The future work on the one hand deals with the study itself, which would mainly focus on a long-term check of how the agile practices influence the daily development. This would mean another assessment (if necessary including a retrospective) after some more months of using the new way of developing. Furthermore, we could address some of the mentioned limitations regarding the generalizability (cf. Section 5), e.g. performing similar studies with different companies from different domains. If we would have had more time for this overall improvement, we would even have implemented the single agile practices after each other.

Regarding the overall initiative, our idea would be to combine the elicited findings, what and how the single agile practices improve, with other existing data to come up with some kind of knowledge for goal-oriented process improvement using agile practices. Finally, another aspect for future work might be the integration of aspects from the agile development (e.g. agile methods or practice) into the used standard ISO 29110. However, this is a generic lack which research is currently trying to overcome, e.g. in the medical domain by creating a technical report [27].

## 7 Acknowlegements

We would like to thank all interview participants for their time and the openness. Furthermore, we thank Prof. Oberweis and Mr. Zehler for providing feedback on the paper.